\documentclass[aps,pre,showpacs,superscriptaddress,groupedaddress]{revtex4}  

\usepackage{graphicx}  
\usepackage{dcolumn}   
\usepackage{bm}        
\usepackage{amssymb}   
\usepackage[usenames]{color}

\begin{document}

\title{Exactly solvable tight-binding model on the RAN: 
fractal energy spectrum and Bose-Einstein condensation}


\author{Maurizio Serva}

\affiliation{Departamento de Biof\'isica e Farmacologia, 
Universidade Federal do Rio Grande do Norte, 59072-970, Natal-RN, Brazil}

\affiliation{Dipartimento di Ingegneria e Scienze dell'Informazione e Matematica, 
Universit\`a dell'Aquila, 67010 L'Aquila, Italy}

\date{\today}

\begin{abstract}
We initially consider a single-particle tight-binding model on the 
Regularized Apollonian Network (RAN). 
The RAN is defined starting from a tetrahedral structure with four nodes all connected (generation 0). At any successive generations,
new nodes are added and connected with the surrounding three nodes.
As a result,  a power-law cumulative distribution of connectivity
$P(k)\propto {1}/{k^{\eta}}$ with $\eta=\ln(3)/\ln(2) \approx 1.585$ is obtained.

The eigenvalues of the Hamiltonian are exactly computed by a 
recursive approach for any size of the network.
In the infinite size limit, the density of states and the cumulative 
distribution of states (integrated density of states) are also exactly 
determined.
The relevant scaling behavior of the cumulative distribution
close to the band bottom is shown to be power law with an exponent 
depending on the spectral dimension and not on the embedding dimension.

We then consider a gas made by an infinite number of non-interacting bosons
each of them described by the tight-binding Hamiltonian on the RAN
and we prove that, for sufficiently
large bosonic density and sufficiently small temperature,
a macroscopic fraction of the particles 
occupy the lowest single-particle energy state forming the 
Bose-Einstein condensate.
We determine no only the transition temperature as a function of
the bosonic density, 
but also the fraction of condensed particle, the fugacity, 
the energy and the specific heat for any temperature and bosonic density.

\end{abstract}


\maketitle

\section{Introduction}

The Bose-Einstein condensation (BEC) is one of the most intriguing
quantum phenomena involving the
formation of a collective quantum state by a gas of identical
non-interacting or weakly-interacting bosons.
Below transition temperature, the collective state emerges as a 
consequence of the fact that a macroscopic fraction of the particles 
occupy the lowest single-particle energy state.
For many decades BEC has remained in the realm of theoretical 
predictions. Only recently it received an impressive
experimental demonstration which, in turn, 
has stimulated a new wealth of theoretical work.

The experimental realization of Bose-Einstein condensate 
was first obtained for weakly-interacting 
low-temperature atoms in a magnetic trap 
\cite{AEMWC1995,DMADDKK1995}
proving that BEC is a purely quantum phenomenon that can take
place even in absence of inter-particle interactions.
More recently, BEC has also been reported in solid state
quasi-particles systems such as excitons, antiferro and
ferromagnetic magnons, even at room temperature
\cite{BLIGC2002,K2006,DDDMSHS2006,GRT2008,PSMSM2013}.

The collective state emerges if the particle
density is sufficiently high and the temperature is 
sufficiently low. 
Dimension is also important;  in fact, it is well known
that free bosons hopping on translationally invariant 
networks cannot undergo Bose-Einstein condensation if the space 
dimension $d$ is less than or equal to 2.
Recent studies suggest the possibility that the network topology may 
act as a catalyst for Bose-Einstein condensation 
allowing condensation even if $d \le 2$.
The presence of BEC has been, in fact, proven, for 
simple $d \le 2$ non-translationally invariant networks as comb, star and wheel 
lattices \cite{BCMRRSV2000,BCRSV2001,BBCV2002,BGMST2004,STSRR2006,VLL2011},
as well as in complex networks as the Apollonian one 
\cite{CAS2008,0MLAA2009,0MLAA2010,OSMLS2013}.

To add more evidence to the catalyst role of topology, 
we shall investigate in this work the properties
of non-interacting bosons hopping on the Regularized Apollonian Network, 
showing that this topology implies macroscopic occupation of the ground 
state at low temperatures.

The RAN \cite{SFA2013,SFA2014}, as well its original 
unregularized version (AN) \cite{AHAS2005},
has a complex architecture characterized by a 
power-law distributed connectivity (scale free) and an average minimal
path between two nodes smaller than any power
of the system size (small-world effect).
These two properties, which are shared by most 
of real world networks, make Apollonian Networks 
an intriguing substrate
for statistical models of physical and biological phenomena,
as BEC \cite{CAS2008,0MLAA2009,0MLAA2010,OSMLS2013}, 
the spreading of epidemics \cite{SCSMFA2013,SCCMSFA2013}, 
the democratic majority vote \cite{LMA2012},
Ising systems \cite{AH2005,AAH2009,SFA2013,SFA2014}, Potts systems 
\cite{AAH2010}, disk packing \cite{DM2005} and even gossip spreading
\cite{LSAH2007}.

In this paper we consider a tight-binding model on the RAN, whose
Hamiltonian has non-vanishing hopping energies only between connected nodes.
The Hamiltonian eigenvalues are achieved by a recursive approach
for any size of the network.
The regularization of the network is essential since it removes
the asymmetry associated to the corners, 
allowing for exact iterative calculations which 
can be done only approximatively using AN \cite{OSMLS2013}.
In the infinite size limit, the density of states and the cumulative 
distribution of states (integrated density of states) are also exactly 
determined.
The relevant scaling behavior of the cumulative distribution
close to the band bottom is shown to be power law, with an exponent 
depending on the spectral dimension and not on the embedding dimension.

We then consider a gas made by an infinite number of non-interacting bosons
each of them described by the tight-binding Hamiltonian on the RAN
and we prove that for sufficiently
large bosonic density and sufficiently small temperature
a macroscopic fraction of the particles 
occupy the lowest single-particle energy state forming the 
Bose-Einstein condensate.
We determine not only the transition temperature as a function of
the bosonic density, but also the fraction of condensed particles, the fugacity, 
the energy and the specific heat for any temperature and bosonic density.

The paper is organized as follows:
in section 2 we describe the RAN model, together with its main properties,
including those which differ with respect to AN and which are relevant
for exact computing. 
Then, in section 3, we define the tight-binding Hamiltonian of 
the single particle and we exactly compute its eigenvalues for any finite 
size of the network.
The density of state for the infinite size network, as well as the 
cumulative distribution of states are computed in section 4, were we also  
describe the properties of the cumulative distribution in the 
relevant region close to the band bottom.
Section 5 is devoted to the description of the thermodynamics of
a system composed by an infinite number of identical bosons, proving
the phase-transition occurrence and finding the transition line in 
the density/temperature plane.
Moreover we determine the fraction of condensed particle, the fugacity, 
the energy and the specific heat for any temperature and
bosonic density. 
Finally, our main achievements are summarized in section 6.

\section{Regularized Apollonian Network (RAN)}
The geometrical construction of the Regularized Apollonian Network 
(RAN) can be done iteratively by initially taking 
a $g=0$ generation network with 4 nodes all connected,
forming a tetrahedral structure with 6 bonds (Fig. 1, left picture).
Each of the 4 triples of nodes individuates a different 
elementary triangle.
At generation $g=1$ a new node is added inside each of
the 4 elementary triangles and it is connected with
the surrounding 3 nodes, so that the network has 8 nodes, 18 bonds
and 12 elementary triangles (Fig. 1, right picture). 
Then the procedure is iterated at any successive generation
inserting new nodes in the last created elementary triangles,
and connecting each of them with the three surrounding nodes.
As a result, one has the following properties for 
a network of generation $g$:

\begin{itemize}
\item[$\bullet$]
the number of new nodes created at any generation $g'$ with
$1 \le g' \le g \,$ is $V_{g'} = 4 \times 3^{g'-1}$;
\item[$\bullet$] 
the total number of nodes at generation $g \ge 0$ is 
$N_{g}=4+\sum_{g'=1}^{g}4 \times 3^{g'-1}=2 \times 3^{g}+2$.
\end{itemize}

The following asymptotic relations holds for large value of g:
the total number of nodes is $N_{g} \simeq 2 \times 3^{g}
\simeq 3  \, N_{g-1}$;
the number of new nodes created at generation $g$ is 
$V_g \simeq (2/3) \, N_{g}$
(the symbol $\simeq$ indicates in this paper that the ratio of the 
two sides tends to 1, 
the symbol $\sim$ indicates the weaker statement that the ratio 
tends to a strictly positive constant, finally, the symbol
$\approx$ will be used for generic approximate equalities).

The connectivity of a node is defined as the number of connections to
other nodes.
In $RAN$ the connectivity of each of the
already existing nodes (the so-called old nodes) is doubled when generation 
is updated, while the connectivity of the
newly created nodes (the new nodes) always equals 3, leading to
the following relevant property: 
\begin{itemize}
\item[$\bullet$]
the connectivity at generation $g$ of a node $i$ only depends on its age. 
More explicitly, its connectivity  is $3 \times 2^{g-g_i'}$ where $g_i'$ 
is the generation at which it was created, and $n_i=g-g_i'$ is its age.
Also, one half of the connections of each old node 
are with new nodes.
\end{itemize}

This property, which is crucial for our exact solution
via renormalization, is not shared by AN. In fact, in AN the 
connectivity depends both on age and geometry since the three 
nodes at the three external corners have a variant connectivity.

The characterization of the RAN can be completed by a 
description of the connectivity distribution, which can be 
straightforwardly deduced by the three previously listed properties:
\begin{itemize}
\item[$\bullet$]
the number of nodes having connectivity $k$ is $m(k,g)$ which 
equals $\,4 \times 3^{g-g'-1}\,$ if $k=3 \times 2^{g'}$ with 
$g'=0,....,g-1$, equals 4 if  $k=3 \times 2^{g}$,
and equals $0$ otherwise.
\end{itemize}
Accordingly, the cumulative distribution of connectivity $P(k)$
is:
\begin{itemize}
\item[$\bullet$]
$P(k)=\sum_{r'\geq k} m(k',g)/N_{g}$ which, for large values of
$g$, exhibits a power-law behavior i.e., $P(k)\propto {1}/{k^{\eta}}$,
with $\eta=\ln(3)/\ln(2) \approx 1.585$,
\end{itemize}
from which the average connectivity can be simply computed
and it is $\simeq 6$, where the approximation holds for large $g$.

The power law cumulative distribution of connectivity $P(k)$ is shared with the AN as well.
Analogously to AN, RAN is scale-free and, as already mentioned, 
it displays the small word effect.
\begin{figure}[!ht]
  \includegraphics[width=6truein,height=4.0truein,angle=0]{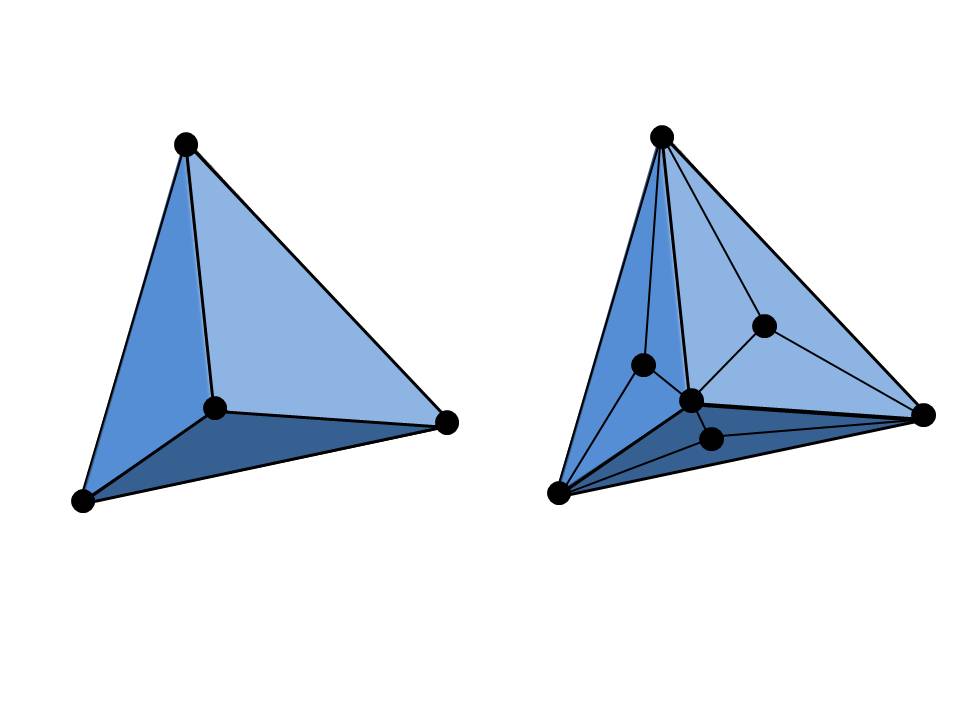}
  \caption{RAN at generation $g=0$ has 4 nodes all connected,
forming a tetrahedral structure with 6 bonds (left picture).
Each of the 4 triples of nodes individuates a different 
elementary triangle.
At generation $g=1$ a new node is added inside each of
the 4 elementary triangles, and it is connected with
the surrounding 3 nodes, so that the network has 8 nodes, 18 bonds
and 12 elementary triangles (right picture). 
The procedure is then iterated at any successive generation
inserting new nodes in the last created elementary triangles,
and connecting each of them with the three surrounding ones.}
  \label{fig1}
\end{figure}
%

We stress once more that the reason for regularizing is that in RAN, 
at variance with AN, the connectivity of the nodes only depends 
on their age.
The removal of the asymmetry associated to the corners 
allows for exact iterative calculations, while in AN
they can be done only approximatively \cite{OSMLS2013}.
\section{Tight-binding Hamiltonian on the RAN}
\subsection{The model}
We consider a tight-binding single-particle Hamiltonian 
on the RAN network. Each node of the
network will be considered as having a single orbital,
whose on node energy will be assumed vanishing without loss
of generality. Moreover, only directly connected nodes
hopping amplitudes are included.
We represent with $| i\rangle$ the state where the particle is 
localized at node $i$. Then, the hopping amplitudes $h_{i,j}$ are
non-null only between connected nodes of the network;
as they are real one has $h_{i,j}=h_{j,i}$. 
Accordingly, the Hamiltonian assumes the form
\begin{equation}
H_g=\sum_{i,j} h_{i,j} | i\rangle \langle j| \, ,
\label{Hg}
\end{equation}
where the sum goes on all the (ordered) connected pairs $i,j$ 
of nodes of RAN of generation $g$. 

The most general class of models
which comply with the symmetry of RAN is achieved assuming that
the hopping amplitudes $h_{ij}$ depend on the connectivities 
of the involved nodes. 
In other words $h_{i,j} = h(k_i, k_j)$. 
Since $k_i = 3 \times 2^{n_i}$, 
where $n_i= g_i-g_i'$ is the age of node $i$, in RAN,
the connectivity dependence is the same as age dependence.
This implies that we can also rewrite $h_{i,j} = h(n_i, n_j)$.

In this work we will assume that the hopping energies are 
$h_{i,j} = t/\sqrt{k_i k_j}$.
The positive parameter $t$ is the only relevant microscopic energy scale
and, from a computational point of view,
it can be chosen without loss of generality equal to 1 
(for $t \neq 1$, all eigenvalues are simply multiplied by $t$).

The off-diagonal Hamiltonian elements $h_{i,j}$
are inversely proportional to the square root
of the product of the connectivity of the two nodes.
This rescaling avoids the divergence of the ground state energy,
while keeping the energy bandwidth finite in the infinite size limit
of the network ($g \to \infty$).
Rescaling of the microscopic energy scale
is commonly used for systems presenting high
connectivity as mean-field models or models with long-range
connections (see for example Refs. \cite{B1982,GKKR1996,AR1995,S2010,S2011})
in order to keep the physical quantities
well defined when the size diverges.

\subsection{Eigenvalues}
In this section we compute the eigenvalues of the Hamiltonian
$H_g$  for any $g$.
The number of eigenvalues equals the number of nodes and, 
therefore, it is $N_g=2 \times 3^{g}+2$.
 
Consider a RAN of generation $g$ on this network and
consider an old node (a node
that is not newly created at last generation $g$). Let us
call this node 0 and let us consider all its $k_0$ connected
nodes labeled by $i$   $\; (i=1,2,...,k_0= 3 \times 2^{n_0}$ where $n_0$ 
is the age of node 0); 
the local Schr\"odinger equation centered over node
0, correspondingly to an eigenvalue $\epsilon$, reads
\begin{equation}
\epsilon \psi_0= 
\sum_{i=1}^{k_0} h_{0,i}\psi_i \, ,
\label{Sg0}
\end{equation}
where $k_0$ is even and
among the $k_0$ connected nodes there are $k_0/2$ newly
created nodes at generation $g$ and $k_0/2$ older nodes.
Let us use the even index  $2i = 2, 4, ...., k_0$  for the newly 
created and an odd index $2i-1= 1, 3, ...., k_0-1$ for the others.

Let us now consider the  $k_0/2$  local Schr\"odinger equations
centered over newly created nodes connected to node 0. We have
\begin{equation}
\epsilon \psi_{2i}=  h_{2i,0}\psi_0 + h_{2i,2i-1}\psi_{2i-1}
+ h_{2i,2i+1}\psi_{2i+1} \, ,
\label{Sgi}
\end{equation}
where we use the convention $\psi_{k_0+1} = \psi_{1}$
and $h_{k_0,k_0+1} = h_{k_0,1}$.

Assume that $\epsilon \neq 0$. In this case (and only in this case)
the $\psi_{2i}$ can be integrated, i.e. they can be substituted
from equation (\ref{Sgi}) into equation (\ref{Sg0}). In this way, we
obtain an equation which connects node 0 with nodes with
an odd index, i.e. an equation which only concerns
those $k_0/2$ older nodes already existing at generation $g-1$.
Then, taking into account that all $h_{0,2i}$
are equal and $h_{0,2i}=h_{2i,0}=1/\sqrt{3 k_0}$,
and considering also that $h_{0,2i}h_{2i,2i-1}=\frac{1}{3}h_{0,2i-1}$
we obtain
\begin{equation}
\left(\epsilon -\frac{1}{6\epsilon} \right)\psi_0=  
\sum_{i=1}^{k_0/2} \left(1 + \frac{2}{3\epsilon} \right) h_{0,2i-1}\psi_{2i-1} \, .
\label{Sg02}
\end{equation}
The above equation, provided that $\epsilon \neq 0$, holds at
any of the nodes which were already present at generation
$g-1$ and, therefore, it can be compared with the local
Schr\"odinger equation centered on node 0 at generation $g-1$
\begin{equation}
\epsilon' \psi_0=  
\sum_{i=1}^{k_0/2} h'_{0,2i-1}\psi_{2i-1} \, .
\label{Sg03}
\end{equation}
Since at generation $g-1$ all connectivities
are one half of connectivities of generation $g$,
the hopping energies $h'_{0,2i-1}$ at generation
$g-1$ are twice the hopping energies $h_{0,2i-1}$ of generation
$g$. Then the two equations coincide provided that
\begin{equation}
2\left(\epsilon -\frac{1}{6\epsilon} \right)=  
\epsilon' \left(1 + \frac{2}{3\epsilon} \right) \, ,
\label{E}
\end{equation}
which proves the following:
a non vanishing $\epsilon$ is an eigenvalue of the Hamiltonian $H_g$ 
if and only if $\epsilon'$ is an eigenvalue of 
the Hamiltonian $H_{g-1}$.

Eq. (\ref{E}) can be solved with respect to $\epsilon$ giving immediately:
\begin{equation}
\epsilon_{1,2}=\frac{1}{4}\left(\epsilon' \pm 
\sqrt{(\epsilon')^2 + \frac{8}{3}(2\epsilon'+1)} \right) \, .
\label{E2}
\end{equation}

As we have assumed that $\epsilon \neq 0$, equation (\ref{E2}) 
only gives the non vanishing eigenvalues of generation $g$, once we know the 
eigenvalues (vanishing and non vanishing) of generation $g-1$. 
If there are missing eigenvalues after using (\ref{E2}), 
they must be vanishing eigenvalues.
Therefore, the $N_g$ eigenvalues of the generation $g$ Hamiltonian
can be generated applying (\ref{E2}) to all $N_{g-1}$
eigenvalues of the generation $g-1$ Hamiltonian.
In this way $2N_{g-1}$ eigenvalues are generated
but only those which are not vanishing are retained.
Then vanishing eigenvalues are added in order that the total number
is $N_g$.
 
Since $N_g \ge 2 \, N_{g-1}$ (the equality only holds when $g=1$ while 
for large $g$ one has $N_g \simeq 3  \, N_{g-1}$),
we can simplify as follows:
\begin{itemize}

\item[$(i)$]
equation (\ref{E2}) is applied to all $N_{g-1}=2 \times 3^{g-1}+2$
eigenvalues of the generation $g-1$ Hamiltonian.
In this way $2N_{g-1}=4 \times 3^{g-1}+4$ eigenvalues
are generated and all retained (including those which are null),
\smallskip 

\item[$(ii)$]
$N_g-2N_{g-1}=2 \times 3^{g-1}-2$ vanishing eigenvalues are added
in order that the total number is $N_{g}=2 \times 3^{g}+2$.  
\end{itemize}

When $g=0$, the RAN reduces to four nodes which are all connected
so that the connectivity equals 3 for all nodes and
all the hopping amplitudes equal $1/3$. Thus, the Hamiltonian 
(\ref{Hg}) can be written as a matrix, yielding
$$
H_0  = \frac{1}{3}
\left[
\begin{array}{cccc}
0 & 1 & 1 & 1 \\ 
1 & 0 & 1 & 1 \\
1 & 1 & 0 & 1 \\
1 & 1 & 1 & 0
\end{array}
\right] \, ,
$$
which has eigenvalues $1,-1/3,-1/3,-1/3$.
Then, the eigenvalues of $H_g$ can be iteratively obtained 
applying the rules (i) and (ii).
 Since for large $g$ the number of vanishing
eigenvalues is $N_g-2N_{g-1} \simeq N_{g}/3$,
about $1/3$ of vanishing are added at any generation updating.

We remark that equation (\ref{E2}) maps real numbers 
in the interval  $[-1/2, 1] $ into
real numbers in the same interval. Therefore, since the
the four eigenvalues corresponding to  $g = 0 $ are in this
range, and at any generation only vanishing eigenvalues 
can be added, the density of states must have
a support in  $[-1/2, 1 $].
It can be also easily verified that at any generation $g>0$
both a single maximum eigenvalue $\epsilon_{max}=1$ and a single
minimum eigenvalue $\epsilon_{0}=-1/2$ are present.

We also remark that $\epsilon_{1,2}$ are always distinct since 
the argument of the square root in (\ref{E2})
never vanishes in the region $[-1/2, 1 $].
On the other hand, reversing  (\ref{E2}) we have that 
$\epsilon'=(6\epsilon^2-1)/(3\epsilon+2)$ so that the same $\epsilon$
cannot be associated to two ore more different $\epsilon'$.
Moreover,  one vanishing eigenvalue $\epsilon$ is created
by (\ref{E2}) from $\epsilon'=-1/2$ excluding the first iteration
from generation $g=0$ to generation $g=1$.
Finally, the number of distinct eigenvalues at generation $g=0$
is 2 and at generation $g=1$ is 4.
In conclusion, the number $L_{g+1}$ of distinct eigenvalues at 
generation $g+1$ exactly satisfies $L_{g+1} = 2 \times L_g $.
 
Summing up, $L_g = 2^{g+1} \sim N_g^{\ln(2)/\ln(3)}$
where $N_g = 2 \times 3^{g}+2$ is the total number of eigenvalues so that
the fractal dimension of the support of the 
density of states is $\nu=1/\eta=\ln(2)/\ln(3) \approx 0.631$
($\eta$ is the exponent in the cumulative distribution of connectivity).

Before concluding this section we would like to stress that analogous calculations 
were already made in \cite{OSMLS2013} for AN but
only approximately or suitably modifying the Hamiltonian. 
On the contrary, all calculations in this section using RAN, 
are exact as well those contained in next sections.

\section{Density of states in the infinite size limit.}
We have seen that for large $g$ the number of eigenvalues
of generation $g-1$ is about $N_{g-1} \simeq 2 \times 3^{g-1}$, while the 
number of eigenvalues 
of generation $g$ is about $N_{g} \simeq 2 \times 3^{g} \simeq 3 N_{g-1}$.
The iterating procedure produces about $2 N_{g-1}$
non-vanishing eigenvalues of generation $g$ from the eigenvalues 
of generation 
$g-1$ by (\ref{iter}). Therefore, about $N_{g-1}$ vanishing 
eigenvalues must be added
in order to complete the list of the $N_{g}  \simeq 3 N_{g-1}$
eigenvalues of generation $g$.

For large $g$, the weight of of vanishing eigenvalues is 
always $1/3$ while, at any iteration of equation (\ref{iter}),  
the weight of any new eigenvalues 
$\epsilon$ is $1/3$ of the weight of the corresponding eigenvalue 
$\epsilon'$. Thus, it is straightforward 
to derive the density of states (\ref{dos}) which holds in the 
infinite size limit $g \to \infty$:

\begin{itemize}

\item[$\bullet$]
Consider the sequence $\bar{\sigma}_k = \sigma_1, \sigma_2,.......,
\sigma_k$ of dichotomous variables $\sigma_i = \pm 1$ and
define iteratively 
\begin{equation}
\epsilon(k+1,\bar{\sigma}_{k+1})= \frac{1}{4} 
\left( \epsilon(k,\bar{\sigma}_{k}) + \sigma_{k+1}
\sqrt{[\epsilon(k,\bar{\sigma}_{k})]^2 +
\frac{8}{3}(2 \, \epsilon(k,\bar{\sigma}_{k}) +1)} \right) \, ,
\label{iter}
\end{equation}
with $ \epsilon(0)=0$.
One has $\epsilon(1,\bar{\sigma}_{1})
= \sigma_1/\sqrt6$ and, in general,
the variables $\epsilon(k,\bar{\sigma})$
only depends on the sequence $\sigma_1, \sigma_2,.......,\sigma_k$
so that for any $k$ they are $2^k$. 

\item[$\bullet$] 
The weight of a given $ \epsilon(k,\bar{\sigma})$ equals 
$\left( 1/3 \right)^{k+1}$ independently on $\bar{\sigma}$.
Then, the density of states in the $ g \to \infty$ limit
is straightforwardly given by
\begin{equation}
\rho(\epsilon)=\sum_{k=0}^\infty  \left( 1/3 \right)^{k+1}
\sum_{\bar{\sigma}_k}   
\delta(\epsilon-\epsilon(k,\bar{\sigma}_k)) \, ,
\label{dos}
\end{equation}
where the notation $\delta(\cdot)$ indicates the Dirac delta function
and the sums go on all possible $k$ and on all possible sequences.
The support of this density is in $[\epsilon_0=-1/2, \epsilon_{max}=1]$.

\end{itemize}

\noindent
Accordingly, the average energy is
\begin{equation}
\int_{\epsilon_0}^1 d \epsilon \rho(\epsilon) \, \epsilon
=\sum_{k=0}^\infty  \left( 1/3 \right)^{k+1}
\sum_{\bar{\sigma}_k} \epsilon(k,\bar{\sigma}_k)=0 \, ,
\label{dos}
\end{equation}
where the second equality can be simply demonstrated using iteratively
(\ref{iter}) with the initial condition $ \epsilon(0)=0$,
which immediately gives 
$\sum_{\bar{\sigma}_k} \epsilon(k,\bar{\sigma}_k)=0$ for any $k$.

The cumulative distribution of states can be 
obtained by the integration of the density of states: 
\begin{equation}
P(\epsilon)= \int_{\epsilon_0}^\epsilon d \epsilon' \rho(\epsilon')=
\sum_{k=0}^\infty  \left( 1/3 \right)^{k+1}
\sum_{\bar{\sigma}_k}  
\theta(\epsilon-\epsilon(k,\bar{\sigma}_k)) \, ,
\label{idos}
\end{equation}
where $\theta(x)$ is the step function
($\theta(x)=0$ for $x < 0$ and $\theta(x)=1$ for $x \ge 0$).
The density of states
$\rho(\epsilon)$ is normalized since, as it is easy to verify, $P(1)=1$.
This cumulative distribution depicted in Fig. 2 is
generated by a $g=20$ network ($N_g = 6,973,568,804$).
The fractal nature of its support is evident,
since there are steps of all sizes.
Moreover, it can be seen that close to extremes 
it presents a self-similar structure
that will be better described in the following.

Let us see more in detail the behavior of the cumulative 
distribution close to the two extremes $\epsilon_{max}=1$ 
and $\epsilon_{0}=-1/2$.
Let us start by rewriting (\ref{idos}) as 
\begin{equation}
P(\epsilon)=\frac{1}{3} \, \theta(\epsilon)+
 \,\sum_{k=0}^\infty \left( 1/3 \right)^{k+2} 
\sum_{\bar{\sigma}_k} \sum_{\sigma_{k+1}} 
\theta(\epsilon-\epsilon(k+1,\bar{\sigma}_{k+1})) \, ,
\label{ideq}
\end{equation}
where $\sigma_{k+1}$ in the last sum
takes the two possible values $\pm 1$.
Using  (\ref{iter}) one gets
\begin{equation}
 \sum_{\sigma_{k+1}} 
\theta(\epsilon-\epsilon(k+1,\bar{\sigma}_{k+1}))
=1+[2\theta(4\epsilon-\epsilon(k,\bar{\sigma}_{k}))-1] \,
\theta(q(\epsilon)-\epsilon(k,\bar{\sigma}_{k})) \, ,
\label{q}
\end{equation}
where $q(\epsilon)=(6\epsilon^2-1)/(3\epsilon+2)$.
When $\epsilon>1/4$, since all
$\epsilon(k,\bar{\sigma}_{k})$ are equal or smaller than unity,
one has $\theta(4\epsilon-\epsilon(k,\bar{\sigma}_{k})) =1$.
Then, for $\epsilon>1/4$, expression ($\ref{q}$)
has the simpler form:
\begin{equation}
 \sum_{\sigma_{k+1}} 
\theta(\epsilon-\epsilon(k+1,\bar{\sigma}_{k+1}))
=1+\theta(q(\epsilon)-\epsilon(k,\bar{\sigma}_{k})) \, ,
\label{q+}
\end{equation}
so that for $\epsilon>1/4$ equation (\ref{ideq}) gives
\begin{equation}
P(\epsilon)=\frac{2}{3} + \frac{1}{3}P(q(\epsilon)) \, .
\label{ideq+}
\end{equation}

For $\epsilon$ close to $\epsilon_{max}=1$, 
using $\epsilon = 1-x$, the linearization of $q(\epsilon)$
gives $q(\epsilon)=q(1-x)\simeq 1-(9/5)x$, then after having defined 
$Q(x)=1-P(1-x)$ we obtain
\begin{equation}
Q(x) \simeq \frac{1}{3}Q(\frac{9}{5}x) \, ,
\label{int+}
\end{equation}
which for any value of $x$
gives rise to sequences of points geometrically spaced 
corresponding to a geometrically valued cumulative distribution.
There are two consequences, the first one being that close to $\epsilon_{max}=1$
the cumulative distribution is self-similar, the second is that 
any of the sequence of points must be on
a geometrical curve $a x^{1+\delta}$ 
which must satisfy (\ref{int+}), which implies
$1+\delta =\ln(3)/\ln(9/5) \approx 1.869$.    
 
When $\epsilon<-1/8$, since all
$\epsilon(k,\bar{\sigma}_{k})$ are equal or larger than $\epsilon_0 = -1/2$,
one has $\theta(4\epsilon-\epsilon(k,\bar{\sigma}_{k})) =0$.
Then, for $\epsilon>1/4$, we have from (\ref{q})
\begin{equation}
 \sum_{\sigma_{k+1}} 
\theta(\epsilon-\epsilon(k+1,\bar{\sigma}_{k+1}))
=1-\theta(q(\epsilon)-\epsilon(k,\bar{\sigma}_{k})) \, ,
\label{q-}
\end{equation}
so that for $\epsilon<-1/8$ equation (\ref{ideq}) gives
\begin{equation}
P(\epsilon)=\frac{1}{3} - \frac{1}{3}P(q(\epsilon)) \, .
\label{ideq-}
\end{equation}

For $\epsilon$ close to $\epsilon_{0}=-1/2$, 
using $\epsilon = x+\epsilon_{0}$, the linearization of 
$q(\epsilon)$ gives $q(x+\epsilon_{0})=1-18x$. 
Therefore 
\begin{equation}
P(x+\epsilon_{0})=\frac{1}{3}[1-P(1-18x)]=\frac{1}{3}Q(18x) \, ,
\label{int-}
\end{equation}
so that the self-similar structure of $Q(x)$
is entirely reflected in $P(x+\epsilon_{0})$,
whose corresponding sequences of points are on geometrical curves
$b (\epsilon-\epsilon_0)^{1+\delta}$.

This can be appreciated in
Fig. 3 which shows the self-similar cumulative distribution 
near the band bottom $\epsilon_0=-1/2$; in particular
it is shown that the two envelopes which 
contain $P(\epsilon)$ and which correspond to two
distinct sequences of points generated by
(\ref{int+},\ref{int-}) behave proportionally to
$(\epsilon-\epsilon_0)^{1+\delta}$
with $1+\delta =\ln(3)/\ln(9/5) \approx 1.869$.
In other words
\begin{equation}b_- (\epsilon-\epsilon_0)^{1+\delta} \le
 P(\epsilon) \le b_+ (\epsilon-\epsilon_0)^{1+\delta} \, ,
\label{ineq}
\end{equation}
with $b_- \approx 19.0$ and $b_+ \approx 38.6$.

This super-linear behavior of the cumulative distribution of states
close to the band bottom points to a BEC transition at finite temperature,
as it will be confirmed in next section.

The exponent $\delta$ corresponds to the spectral dimension 
$d_s=2 \, (1+ \delta) \approx 3.738$ associated to the generator of 
the process with jump rates $h_{i,j}$ between connected nodes.
Notice that the spectral dimension has a value larger then 2 while
the embedding dimension of the network has dimension 2.

\begin{figure}[!ht]
  \includegraphics[width=5.5truein,height=4.0truein,angle=0]{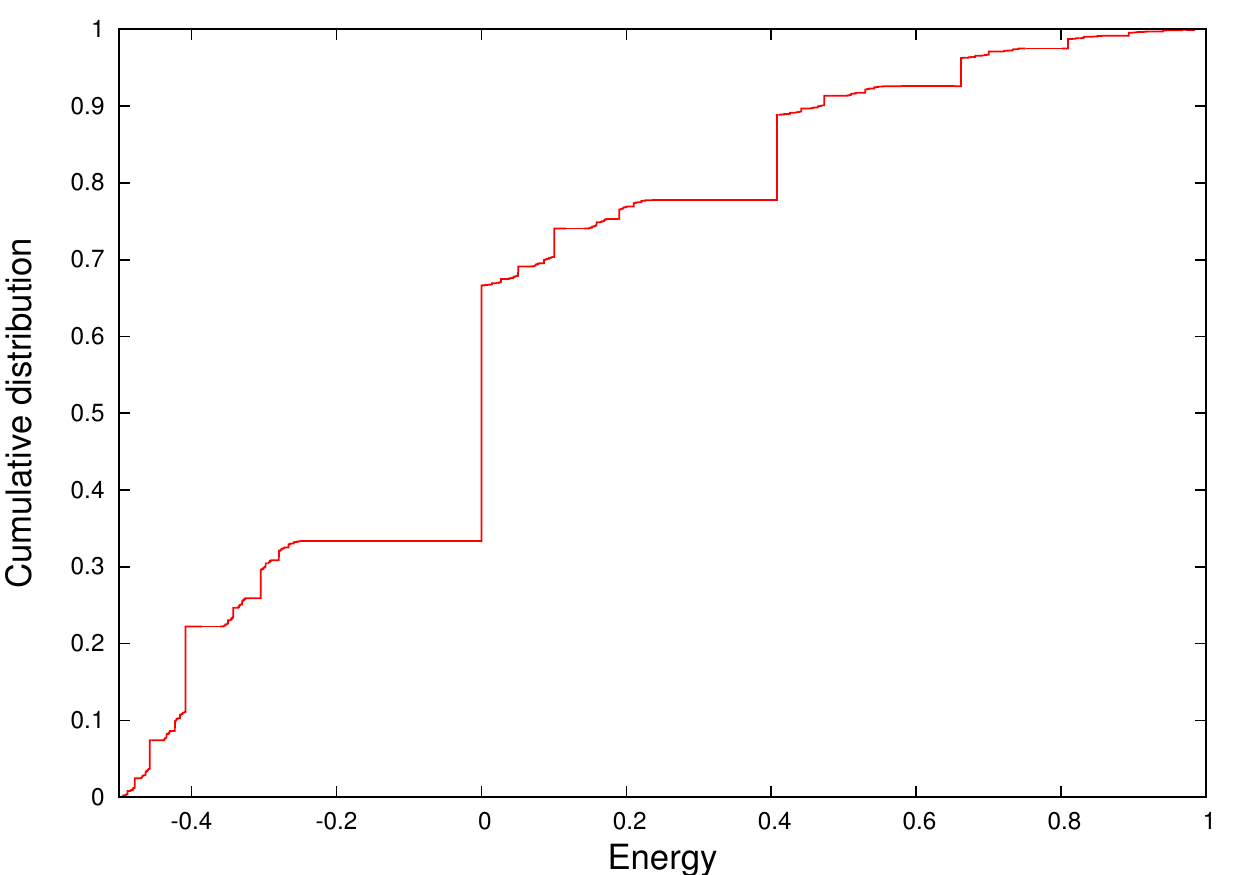}
  \caption{Cumulative distribution of states $P(\epsilon)$.
Degeneracies and gaps are signaled by vertical and horizontal 
segments, respectively. Close to the extremes $P(\epsilon)$
shows a self-similar structure.
This figure has been obtained considering 
a $g=20$ network corresponding to $N_g = 6,973,568,804$ nodes.}
  \label{fig2}
\end{figure}
%
%
\begin{figure}[!ht]
  \includegraphics[width=5.5truein,height=4.0truein,angle=0]{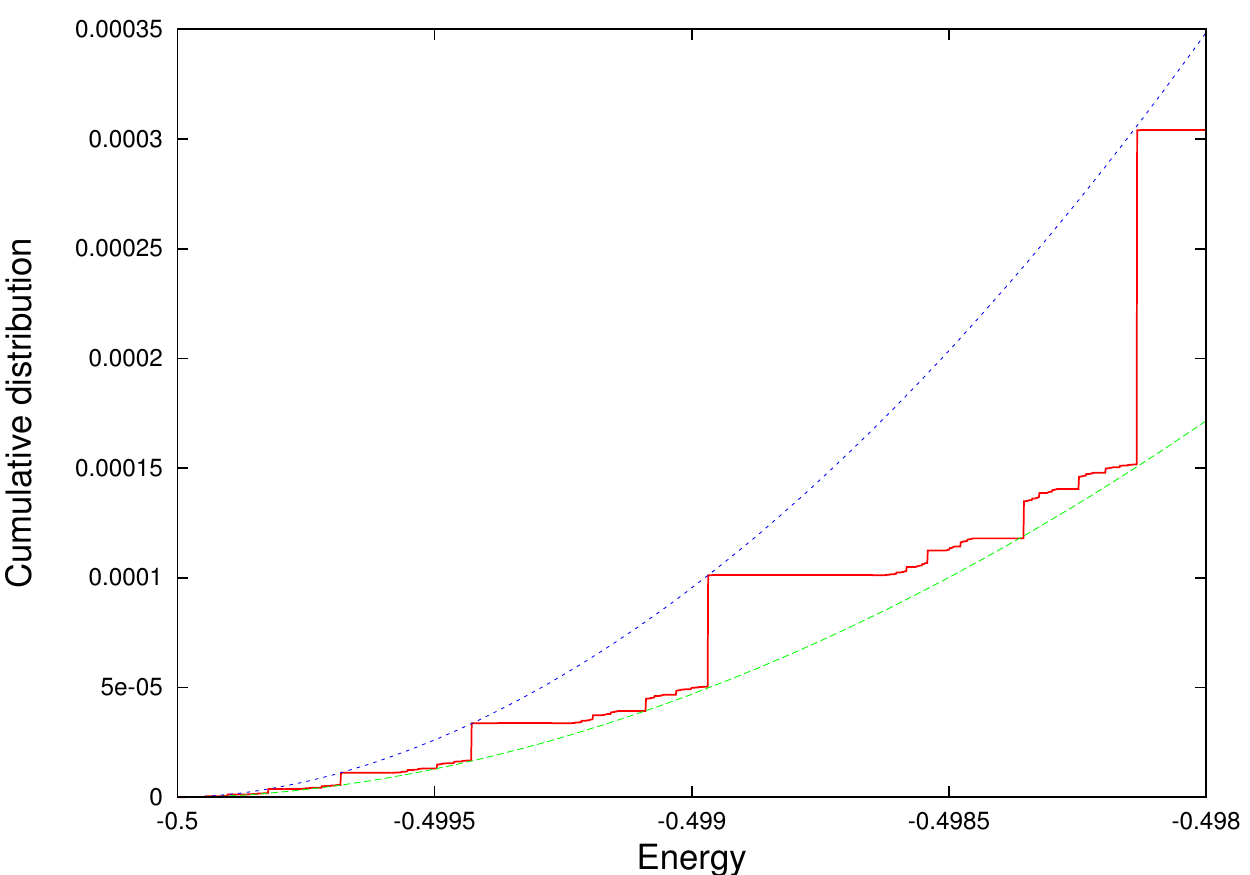}
  \caption{Cumulative distribution of states $P(\epsilon)$
close to the band bottom plotted together with the two envelopes
$b_- \, (\epsilon+0.5)^{1+\delta}$ and
$b_+ \,  (\epsilon+0.5)^{1+\delta}$ 
($b_- \approx 19.0$ and $b_+ \approx 38.6$).
This super-linear behavior of the cumulative distribution
close to the band bottom, corresponding to the spectral 
dimension $d_s=2 \, (1+ \delta) \approx 3.738$,
points to a BEC transition at finite temperature.}
  \label{fig3}
\end{figure}
%

\section{Thermodynamics.}
Let us now consider a gas of $N_p$ non interacting bosons
which can occupy any of the eigenstates corresponding
to the $N_g$ eigenvalues of the Hamiltonian. The average number of 
bosons in each eigenstate is, therefore, $N_p/N_g$
which we name bosonic density.
In the thermodynamic limit $N_p$ diverges while $N_g$ diverges
in the infinite network size limit.
In order to have a non trivial thermodynamics, 
it is compulsory to assume that the bosonic density remains
finite and strictly positive.
In other words, we assume that in the limit where both 
$N_p$ and $N_g$ diverge one has $ N_p/N_g \to \phi$
with $0 < \phi < \infty$. 

We also define $\phi_0$ as the density of condensed bosons
(number of bosons in the ground state divided by the
number of eigenstates) and $\phi_{th}$ as the 
density of thermal bosons (number of bosons not
in the ground state divided by the number of eigenstates).
Then one trivially has $\phi= \phi_0+ \phi_{th}$.

We can then work directly in the thermodynamic limit
(see \cite{BCRSV2001}) and write 
\begin{equation}
\phi= \phi_0+ \phi_{th}= \phi_0+
\int_{\epsilon_0}^1 d \epsilon \, \rho(\epsilon) \, 
\frac{1}{z^{-1}e^{\beta (\epsilon-\epsilon_0)} -1} \, ,
\label{phith}
\end{equation}
where $\beta=1/T$ is the inverse temperature (units 
with Boltzmann constant $K_B =1$) and $z$ is the fugacity.

For $T<T_c$ one has $z=1$ and $\phi_0(\phi, T)= \phi-\phi_{th}(T)$.
Notice that, since $z=1$, $\phi_{th}(T)$ only depends on temperature 
and not on $\phi$.
For $T>T_c$ one has $z<1$ and $\phi_0= 0$
which implies $\phi_{th}=  \phi$. When $\phi_0= 0$ 
equation (\ref{phith}) gives $z(\phi, T)$ as implicit solution.

\subsection{Critical temperature.}
The critical temperature $T_c(\phi)=1/\beta_c(\phi)$ 
is found as solution of (\ref{phith}) setting both $z=1$ and $\phi_0= 0$:
\begin{equation}
\phi= \int_{\epsilon_0}^1 d \epsilon \, \rho(\epsilon) \, 
\frac{1}{e^{\beta_c (\epsilon-\epsilon_0)} -1} \, ,
\label{tc}
\end{equation}
which is depicted in Fig. 4 in a log-log plot showing two
different power-law regimes that can be deduced from
(\ref{tc}). 

When $T_c$ is small ($\beta_c$ is large) only values of $\epsilon$ 
close to $\epsilon_0$ contribute to the integral.
Thus, taking into account that for small values of $\epsilon$ one has
$P(\epsilon)=\int_{\epsilon_0}^\epsilon d \epsilon' \, \rho(\epsilon') 
\sim (\epsilon-\epsilon_0)^{1+\delta}$, one gets
\begin{equation}
\phi \sim \, \int_{\epsilon_0}^1 d \epsilon \, 
\frac{(\epsilon-\epsilon_0)^{\delta} }
{e^{\beta_c (\epsilon-\epsilon_0)} -1}=
T_c^{1+\delta}\int_{0}^{\frac{3}{2} \beta_c} d \epsilon \, 
\frac{x^{\delta}}{e^{x} -1} \, ,
\label{tc-1}
\end{equation}
and finally, neglecting terms exponentially small in $\beta_c$,
one gets
\begin{equation}
T_c \simeq  a_- \phi^{\frac{1}{1+\delta}} \, ,
\label{tc-3}
\end{equation}
where $a_-$ is numerically determined
to be $a_- \approx 0.1$ .

On the contrary, when $T_c$ is large ($\beta_c$ is small) one has 
$e^{\beta_c (\epsilon-\epsilon_0)} -1 \simeq 
\beta_c (\epsilon-\epsilon_0)$
and straightforwardly from (\ref{tc}) we derive
\begin{equation}
T_c \simeq a_+ \phi \, ,
\label{tc+}
\end{equation}
where $a_+=1/\int_{\epsilon_0}^1 d \epsilon \, \rho(\epsilon) \, 
\frac{1}{\epsilon-\epsilon_0} \approx 0.15$.
These two regimes are confirmed by plot in Fig.4.

\begin{figure}[!ht]
  \includegraphics[width=5.5truein,height=4.0truein,angle=0]{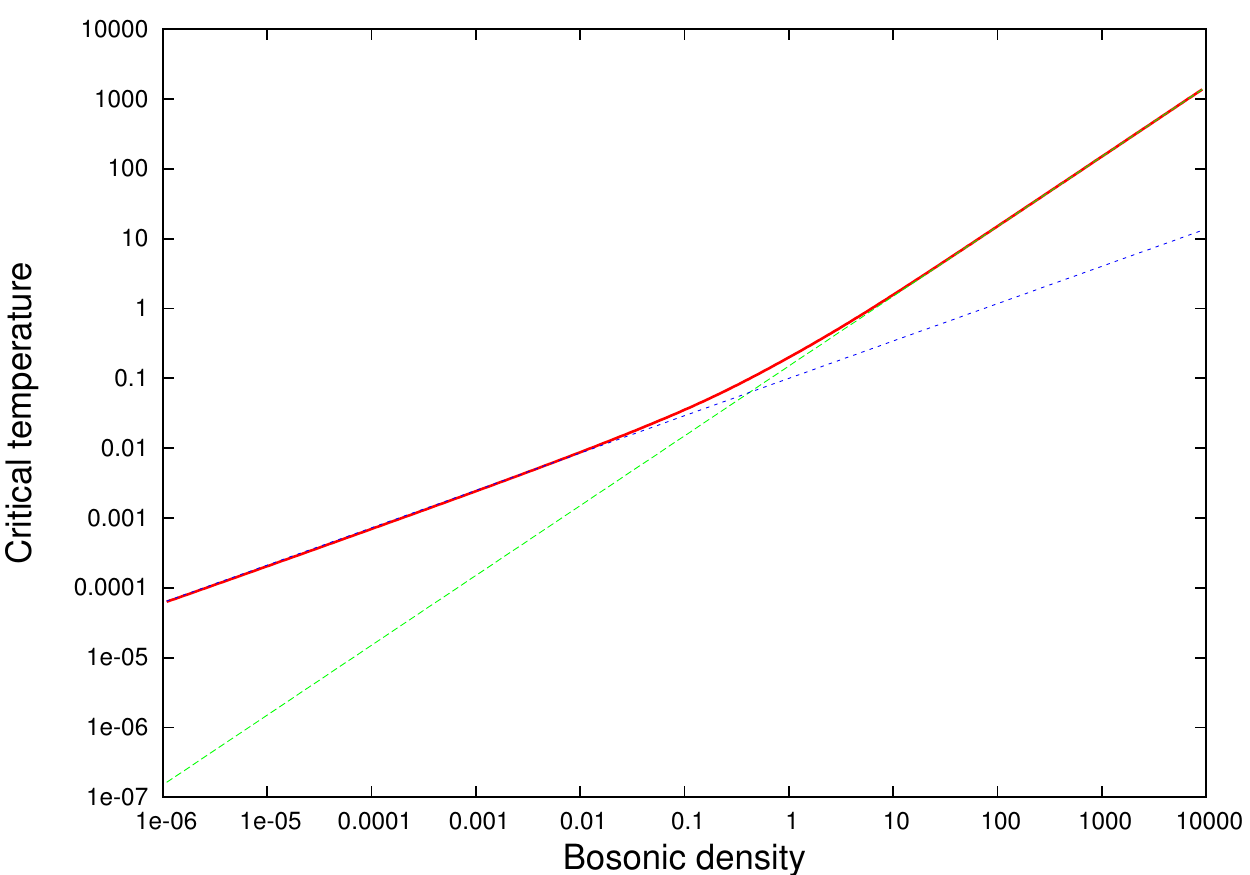}
  \caption{Critical temperature $T_c$ versus bosonic density 
$\phi$ compared with
$0.1 \, \phi^{1/(1+\delta)}$ (small bosonic density) and
$0.15 \, \phi$ (large bosonic density).}
  \label{fig4}
\end{figure}
%

\subsection{Bose-Einstein condensate.}

For $T<T_c$ the condensed fraction $f_0(\phi,T)= \phi_0/\phi$
is
\begin{equation}
f_0(\phi,T) = 1- \frac{\phi_{th}(T)}{\phi}
= 1- \frac{1}{\phi}
\int_{\epsilon_0}^1 d \epsilon \, \rho(\epsilon) \, 
\frac{1}{e^{\beta (\epsilon-\epsilon_0)} -1} \, ,
\label{cf}
\end{equation}
which is depicted in Figs 5, 6 and vanishes for $T=T_c$
where $\phi_{th}= \phi$.

Using exactly the same argument as above one obtains that
for small temperature $T$ (large $\beta$) one has 
$\phi_{th} \sim T^{1+\delta}$ while for large
temperature $T$ (small  $\beta$) one has 
$\phi_{th} \sim T$.
Therefore, since a small $\phi$ implies a small $T_c$
and since $f_0(\phi,T)$ is non vanishing only when  $T<T_c$
one has 
\begin{equation}
f_0(\phi,T) \simeq 1- \left(\frac{T}{T_c} \right)^{1+\delta} \, ,
\label{cf-}
\end{equation}
which holds for small $\phi$ and where $T_c$ depends on $\phi$
according to (\ref{tc}) approximated by (\ref{tc-3})
(see Fig. 5).

On the contrary, a large $\phi$ implies a large $T_c$
and since $f_0(\phi,T)$ is non vanishing in all region $T<T_c$
one has that the previous relation only holds for $T$ close to 0
while for larger $T$ one has
 
\begin{equation}
f_0(\phi,T) \simeq 1- \left(\frac{T}{T_c} \right) \, ,
\label{cf-}
\end{equation}
which holds for large $\phi$ and where
the critical temperature $T_c$ depends on $\phi$
according to (\ref{tc}) approximated by (\ref{tc+})
(see Fig. 6).

Interestingly, for small densities, the behavior is determined by the bottom
of energy band and it corresponds
to a free boson gas in dimension $d_s=2 (1 + \delta) \approx 3.738$
($d_s$ is the spectral dimension).

%
\begin{figure}[!ht]
  \includegraphics[width=5.5truein,height=4.0truein,angle=0]{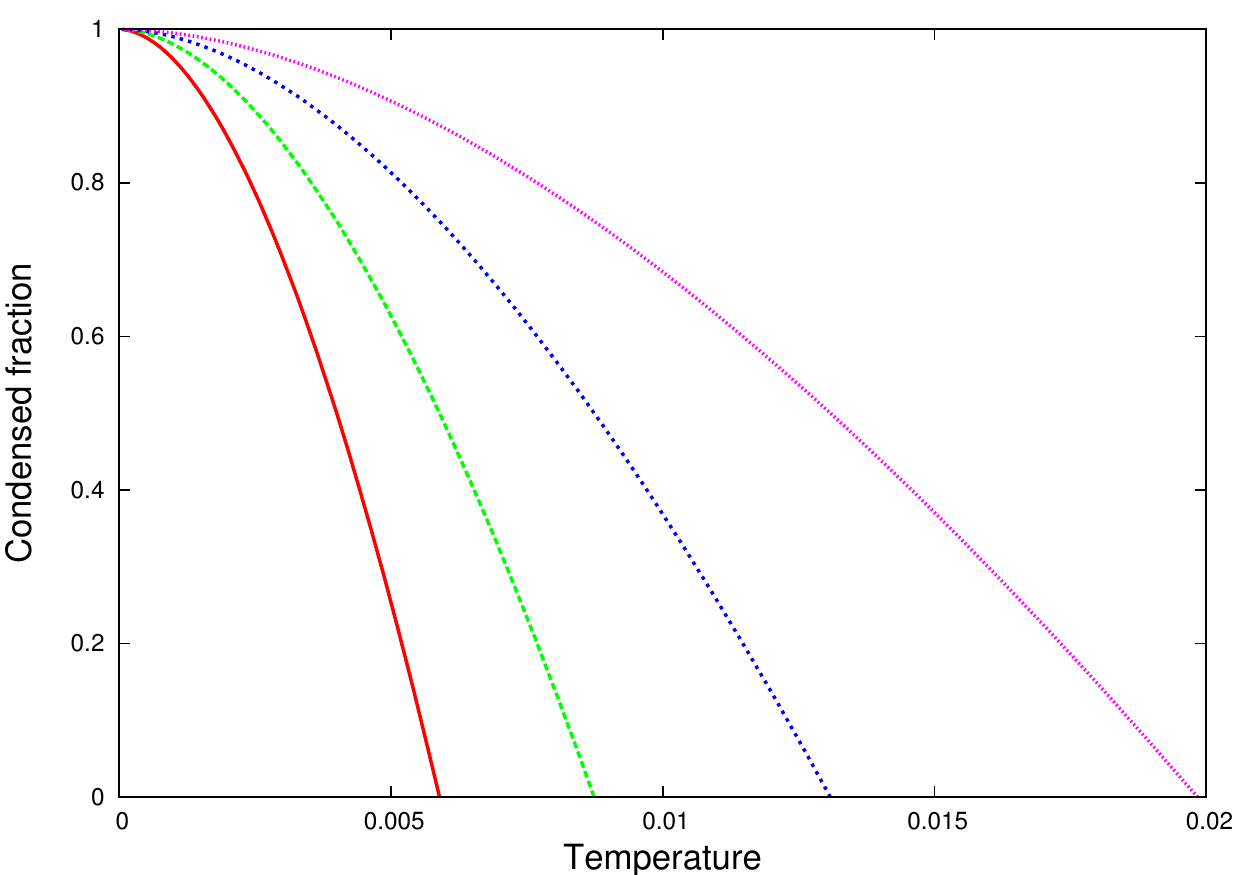}
  \caption{Condensed fraction $f_0(\phi,T)$ versus temperature 
$T$ for $\phi=0.005$, $\phi=0.01$, $\phi=0.02$, $\phi=0.04$
(from left to right).
For these  small densities, the behavior corresponds
to a free boson gas in dimension $d_s=2 (1 + \delta) \approx 3.738$
($d_s$ is the spectral dimension of the model).
One has $f_0(\phi,T) \simeq 1- \left(\frac{T}{T_c} \right)^{1+\delta}$.}
  \label{fig5}
\end{figure}
%
%
\begin{figure}[!ht]
  \includegraphics[width=5.5truein,height=4.0truein,angle=0]{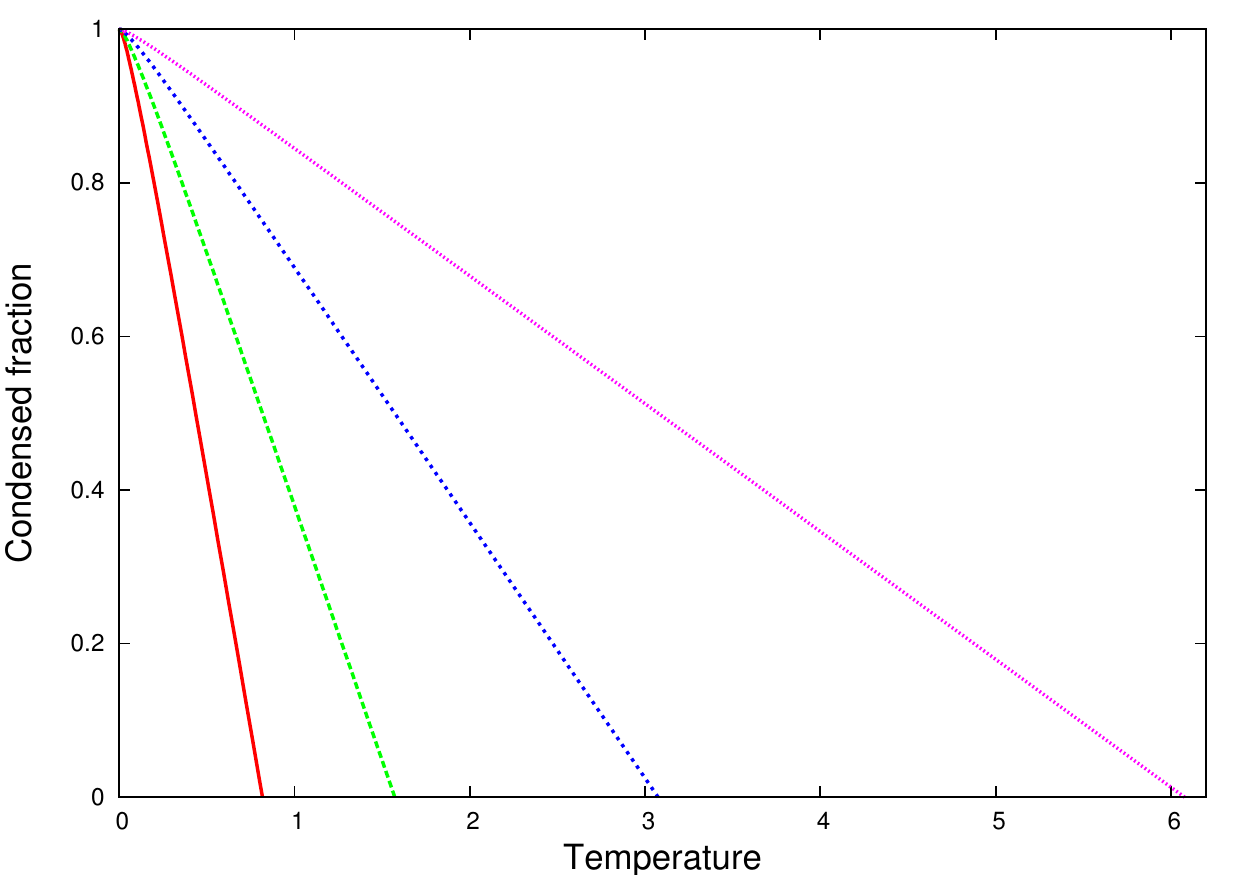}
  \caption{Condensed fraction $f_0(\phi,T)$ versus temperature 
$T$ for $\phi=5$, $\phi=10$, $\phi=20$, $\phi=40$
(from left to right).
For these large densities, the behavior is determined by the fact that 
there is a finite maximum eigenvalue of the Hamiltonian.
One has $f_0(\phi,T) \simeq 1- \left(\frac{T}{T_c} \right)^{}$.}
  \label{fig6}
\end{figure}

\subsection{Fugacity.}
The fugacity $z(\phi, T)$ equals 1 for $T<T_c$
while for $T>T_c$ can be obtained from (\ref{phith})
setting $\phi_0=0$ which gives
\begin{equation} 
\phi= 
\int_{\epsilon_0}^1 d \epsilon \, \rho(\epsilon) \, 
\frac{1}{z^{-1}e^{\beta (\epsilon-\epsilon_0)} -1} \, ,
\label{zeta}
\end{equation}
and, in particular,  when $\beta=0$ $(T \to \infty)$
one has $\phi= 1/(z^{-1}-1)$ which gives
$ \lim_{T \to \infty }z(\phi, T)=\phi/(\phi+1)$.

Then, from (\ref{phith}),
keeping $\phi$ constant and deriving with respect to the 
inverse temperature, one gets
\begin{equation}
\frac{\partial z}{\partial \beta}  = \frac{
z \int_{\epsilon_0}^1 d \epsilon \, \rho(\epsilon) \, 
\frac{(\epsilon-\epsilon_0)e^{\beta (\epsilon-\epsilon_0)}}
{[z^{-1}e^{\beta (\epsilon-\epsilon_0)} -1]^2}
 }{
\int_{\epsilon_0}^1 d \epsilon \, \rho(\epsilon) \, 
\frac{e^{\beta (\epsilon-\epsilon_0)}}
{[z^{-1}e^{\beta (\epsilon-\epsilon_0)} -1]^2} } \, .
\label{derzeta}
\end{equation}

When $z=1$, the integrand in the numerator of right-hand side of
(\ref{derzeta}) behaves as $ \rho(\epsilon) \, (\epsilon-\epsilon_0)^{-1}$
while the integrand in the denominator as
$\rho(\epsilon) \, (\epsilon-\epsilon_0)^{-2}$.
Since $P(\epsilon)= \int_{\epsilon_0}^\epsilon d \epsilon' \, \rho(\epsilon')
\sim (\epsilon-\epsilon_0)^{1+\delta} \approx (\epsilon-\epsilon_0)^{1.869}$,
the numerator is finite while the denominator diverges, 
so that equation (\ref{derzeta}), when $z=1$ (when $T \le T_c$), gives
$\frac{\partial z}{\partial \beta}=0$.
Therefore, equation (\ref{derzeta}) can be used to
operatively compute $z$  for all temperatures 
given that at $\beta=0$ one has $z= \phi/(\phi+1)$.
The result is shown in Fig. 7 where $z(\phi,T)$ is plotted 
against the rescaled temperature $T/T_c$ for various values of $\phi$.

%
\begin{figure}[!ht]
  \includegraphics[width=5.5truein,height=4.0truein,angle=0]{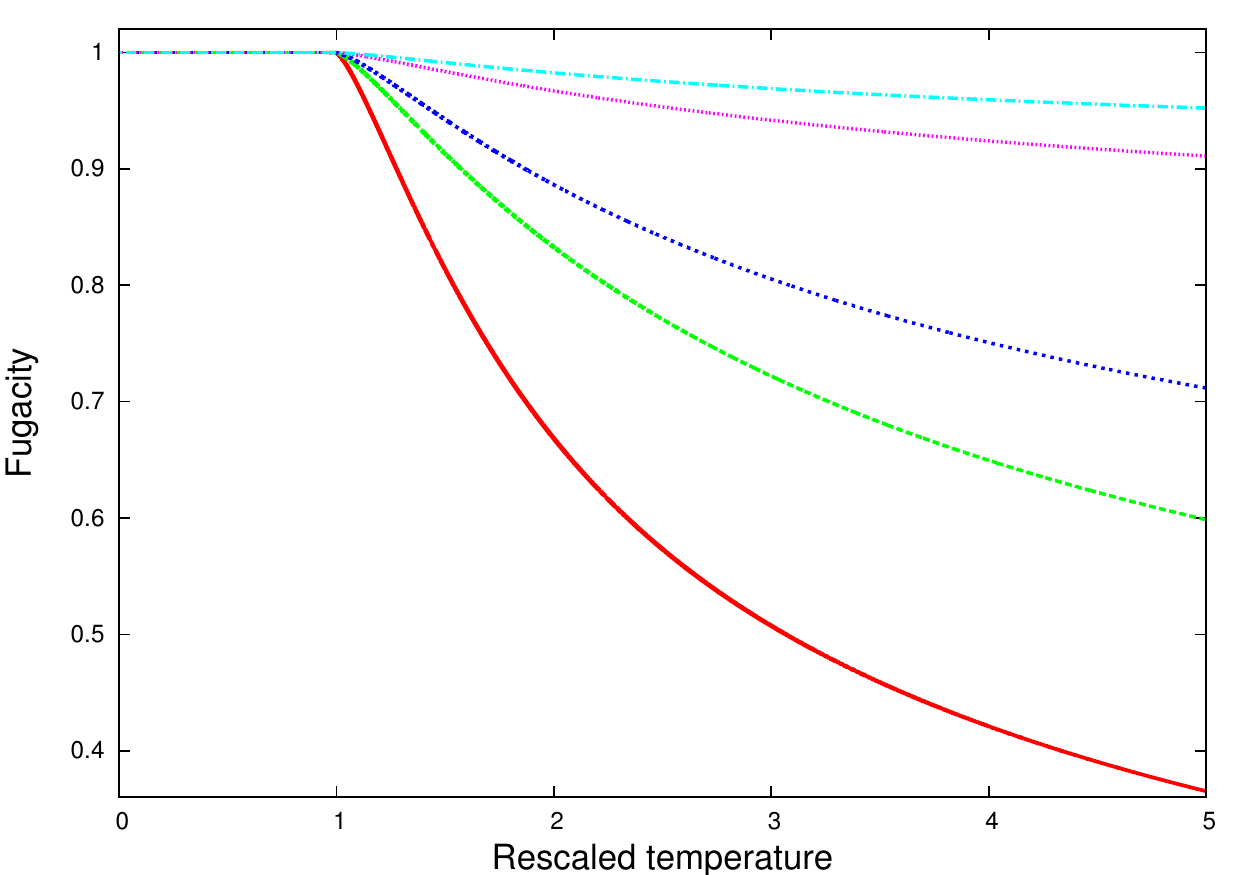}
  \caption{The fugacity $z(\phi,T)$  versus the rescaled temperature 
$T/T_c$ for $\phi=.1$, $\phi=.5$, $\phi=1$, $\phi=5$ and $\phi=10$ 
(from below). The asymptotic value for $T \to \infty$
is $z= \phi/(\phi+1)$. The fugacity behaves linearly in the proximity 
of the transition.}
  \label{fig 7}
\end{figure}
%

\subsection{Energy and specific heat}

The energy density $E(\phi, T)$ (thermodynamic limit 
of the ratio between the total energy and 
number of eigenstates) is given by
\begin{equation} 
E= 
\int_{\epsilon_0}^1 d \epsilon \, \rho(\epsilon) \,  
\frac{\epsilon }{z^{-1}e^{\beta (\epsilon-\epsilon_0)} -1}+
\epsilon_0 \phi_0 =
\int_{\epsilon_0}^1 d \epsilon \, \rho(\epsilon) \,  
\frac{\epsilon -\epsilon_0}{z^{-1}e^{\beta (\epsilon-\epsilon_0)} -1}+
\epsilon_0 \phi \, .
\label{ener}
\end{equation}
In the $T \to \infty$ limit one has that $z = \phi/(1+\phi)$
therefore 
\begin{equation} 
\lim_{T \to \infty} E= 
\phi \int_{\epsilon_0}^1 d \epsilon \, \rho(\epsilon) \,  \epsilon 
= 0 \, ,
\label{elimit}
\end{equation}
where the second equality was proven in Section 4.
On the other hand, in the limit $T \to 0$. as it is obvious, one has
$E \to \epsilon_0 \phi= -\phi/2$.

The specific heat density  $C(\phi, T)$ is obtained by deriving
(\ref{ener}) with respect to the temperature
keeping the bosonic density fixed. 
\begin{equation} 
C = \frac{1}{z^2 T^2}
\int_{\epsilon_0}^1 d \epsilon \, \rho(\epsilon) \, 
\frac{z(\epsilon -\epsilon_0)- \frac{\partial z}{\partial \beta}}
{[z^{-1}e^{\beta (\epsilon-\epsilon_0)} -1]^2}
\, (\epsilon -\epsilon_0) e^{\beta (\epsilon-\epsilon_0)} \, ,
\label{spec}
\end{equation}
where $\frac{\partial z}{\partial \beta}$, given by 
equation (\ref{derzeta}) vanishes for $T \le T_c$.

The specific heat below the critical temperature
is independent on $\phi$ since $z=1$ and 
$ \frac{\partial z}{\partial \beta}=0$.
This fact can be appreciated in Figs. 8 and 9
(notice that the four curves corresponding to different values of $\phi$
are in two different figures because the involved scale of temperature is different).
At the critical temperature the specific heat has a cusp,
with a finite derivative at the left side and infinite derivative at
the right side as it is depicted both in Fig. 8 and in Fig. 9.

%
\begin{figure}[!ht]
  \includegraphics[width=5.5truein,height=4.0truein,angle=0]{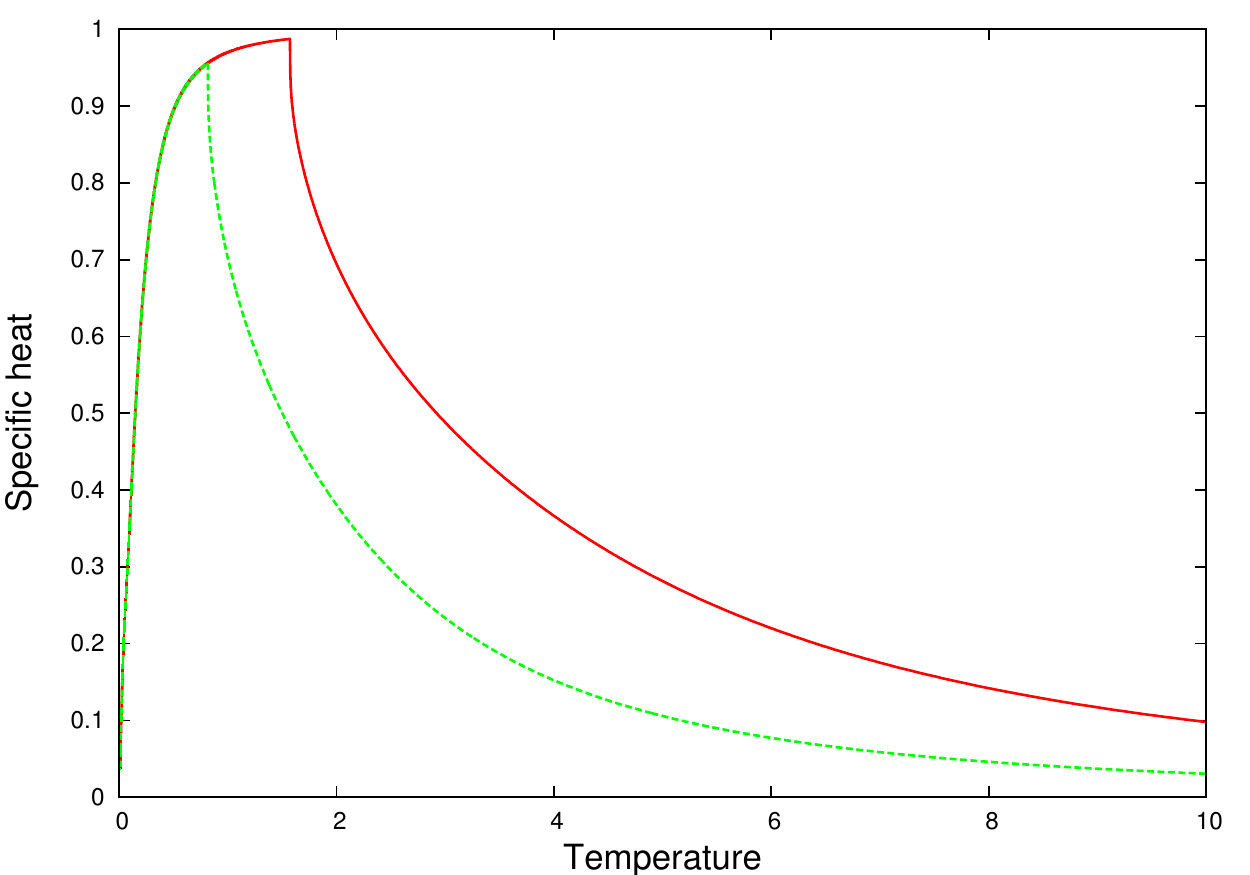}
  \caption{Specific heat $C$ versus temperature 
$T$ for $\phi=5$ (green, dashed line) and $\phi=10$ (red, full line).
The specific heat below the critical temperature
is independent on $\phi$ as it is depicted in the figure.
At the critical temperature $C$ has a cusp,
with a finite derivative at the left side and infinite derivative at
the right side.}
  \label{fig 8}
\end{figure}
%
%
%
\begin{figure}[!ht]
  \includegraphics[width=5.5truein,height=4.0truein,angle=0]{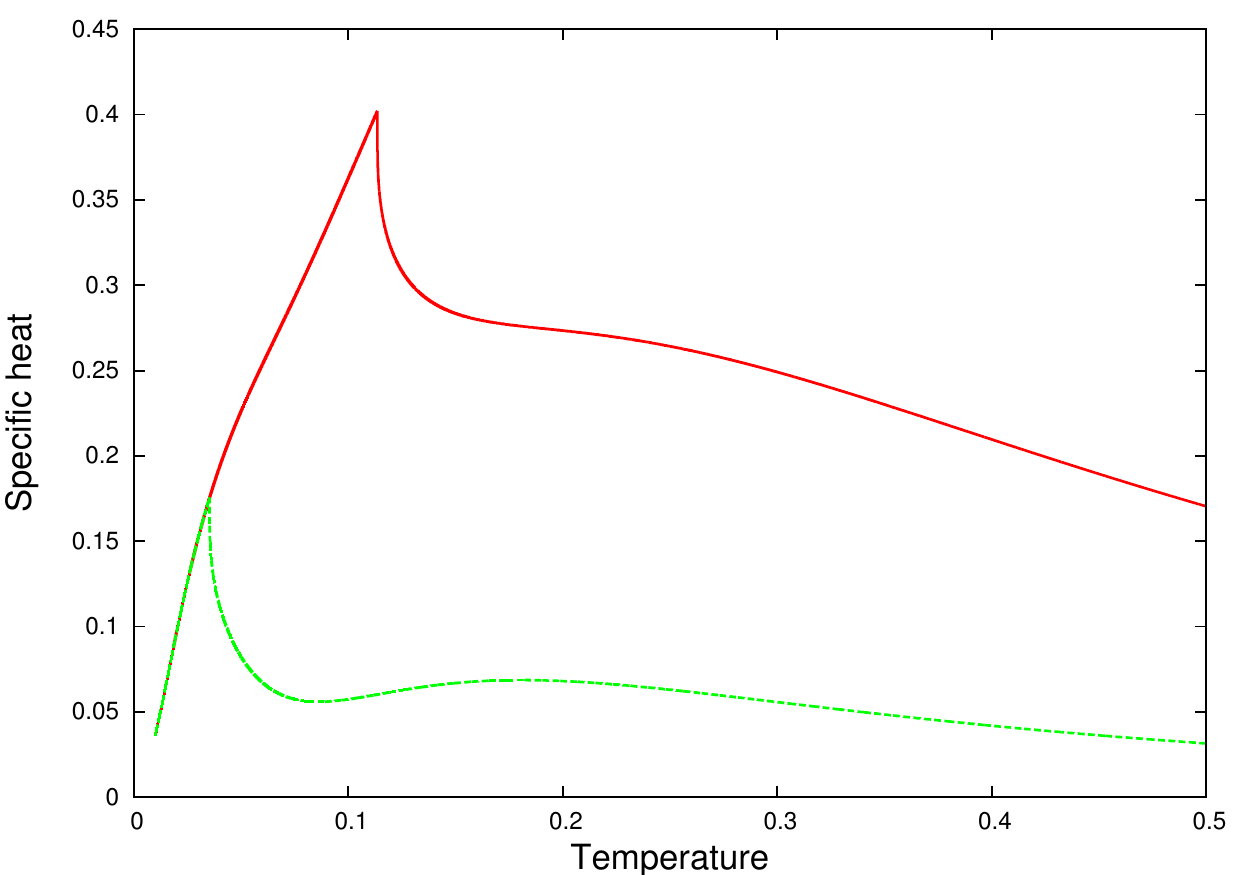}
  \caption{Specific heat $C$ versus temperature 
$T$ for $\phi=.1$ (green, dashed line) and $\phi=.5$ (red, full line).
Notice again that the specific heat below the critical temperature
is independent on $\phi$, while at the critical temperature it has a cusp
with a right side infinite derivative.}
  \label{fig 9}
\end{figure}
%

\section{Discussion}

The main results out of this paper are:

\begin{itemize}
\item[(i)] the exact computation of the eigenvalues 
of a tight-binding Hamiltonian on the RAN substrate 
for any finite size of the network; 
\item[(ii)] the exact computation of the 
cumulative density of states of this 
Hamiltonian in the limit of infinite size network 
for which we also described the properties in the relevant region 
close to the ground state energy;
\item[(iii)] the proof of BEC and its complete description in terms
of transition line, 
fraction of condensed particles, fugacity, energy and specific heat.
\end{itemize}

Since RAN is embedded in a two dimensional space,
this result add more evidence to the catalyst role
of topology for the existence of a Bose-Einstein condensate.
The relevant parameter for the existence of
the condensate is the spectral dimension $d_s$
which has to be larger then 2.
Since the spectral dimension $d_s$ and the physical dimension
$d$ coincide for translationally invariant lattices,
there is not contradiction with the prescription 
$d>2$ for these models.
In this present model, on the contrary, we have $d_s \approx 3.738 >d=2$
which explains how the topology may induce condensation.

\begin{acknowledgements}
The author thanks Eudenilson Lins Albuquerque, Umberto Laino Fulco,
Marcelo Leite Lyra and Pasquale Sodano for discussions, suggestions
and a critical reading of the manuscript.
Financial support from the Brazilian 
Research Agencies CAPES (Rede NanoBioTec and PNPD), CNPq 
[INCT-Nano(Bio) Simes, Casadinho-Procad] and FAPERN/CNPq (PRONEM) and 
PRIN 2009 protocollo n. 2009TA2595.02 is also acknowledged.
\end{acknowledgements}



\end{document}